# Autonomous computational catalysis through an agentic research system

**Authors:** *Honghao Chen* [1], *Jiangjie Qiu* [1], *Yi Shen Tew* [1], *Xiaonan Wang* [1*]

**Affiliations:**

*1.* Beijing Key Laboratory of Artificial Intelligence for Advanced Chemical Engineering Materials, State Key Laboratory of Chemical Engineering and Low-Carbon Technology, Department of Chemical Engineering, Tsinghua University, 100084, Beijing, China.

**Corresponding author:** Xiaonan Wang, email: wangxiaonan@tsinghua.edu.cn

**Abstract**

Autonomous agents are beginning to transform scientific research from tool-assisted workflows toward self-sustaining discovery processes. Computational catalysis provides a representative challenge, as catalyst discovery requires high-level questions to be translated into coordinated model construction, atomistic simulation, mechanistic analysis, and iterative design across multiple scales.

Here we introduce CatMaster, a catalysis-native agentic research system that recasts computational catalysis as a low-barrier virtual ecosystem for autonomous research. CatMaster maintains an evolving research state and extends capabilities through self-feedback across model construction, calculation, critique and catalyst-design decisions within one extensible environment. Across progressively challenging tasks, CatMaster converts natural-language requests into concrete computational studies, from essential atomistic modelling and standard calculations to mechanism exploration and closed-loop catalyst design. It showed robust execution in representative computational-catalysis scenarios and near-leading performance across selected MatBench tasks, with phonons scenario demonstrating its modelling self-evolution capability. In the independent $CO_2$-to-CO catalyst design case, CatMaster used iterative self-critique and evidence refinement to identify competitive B–$CoN_4$ and $NiN_3B$/N–$NiN_3B$ motifs. These results establish a virtual-ecosystem paradigm in which AI agents move beyond simulation execution toward end-to-end computational research, providing a foundation for autonomous discovery in catalysis and materials science.

## Introduction

Heterogeneous catalysis underpins industrial chemistry, sustainable energy and emissions control.[1] Improving activity, selectivity and stability increasingly depends on mechanism-guided design, where catalytic performance is connected to surface structure, reaction energetics and elementary steps. Together with in situ and operando characterization, density functional theory (DFT) and atomistic simulations provide molecular routes for making these connections, linking candidate interfaces to atomistic evidence and design hypotheses.[2-5] Artificial intelligence (AI) has strengthened this atomistic simulation by accelerating property prediction and catalyst screening with learned descriptors and surrogate models, and by extending structure and reaction exploration through machine-learning interatomic potentials.[6-11] By lowering the cost of prediction and exploration, these models broaden the candidate structures, reaction pathways and mechanistic tests available to a catalyst problem while keeping physical validation and chemical interpretation central to catalyst discovery.

As these modelling capabilities multiply, the difficulty increasingly lies in assembling different modelling and simulation tools into a coherent catalyst study. A single catalyst question can now draw on literature-derived structures, descriptor models, machine-learning potentials, DFT calculations, electronic analyses and mechanistic searches. Each layer has a different purpose, cost and domain of validity, and moving between them requires choices about structural models, evidence hierarchy, verification level and mechanistic interpretation. In practice, scientists must translate abstract mechanistic ideas into concrete compatible structures, parameters, compute jobs and chemical checks across fragmented toolchains. This heterogeneity keeps the entry barrier high even as individual models and codes become more powerful.

To manage this heterogeneity across structures, calculations, data and execution, computational chemistry and materials science have developed substantial research infrastructure. Libraries and workflow platforms such as ASE, pymatgen, FireWorks, AiiDA and the Materials Project make atomistic structures, calculations, materials data and execution pipelines reusable across studies.[12-16] Adapting these infrastructures to a new catalysis question still requires predefined workflows, maintained interfaces and substantial user expertise. Catalysis studies are especially demanding because active-site models, surface terminations, adsorbate states, reaction paths and validation levels are often customized for each mechanism or design hypothesis. Recent autonomous-science systems suggest a possible route to this problem. Recent systems connect natural-language scientific intent with executable investigation, moving AI from passive command-driven tools toward active AI-assisted discovery systems by supporting hypothesis generation, literature synthesis, data analysis, empirical software creation.[17-22] Agentic frameworks have also begun to demonstrate in fragmented quantum and solid-state simulation workflows, cross-domain tool use and catalyst-screening studies.[23-32]

These studies establish useful pieces of a new research method, yet they leave the essential need for an evolving research state that keeps model construction, calculations, interpretation and design decisions connected throughout a computational catalysis study unresolved. Most work remains centered on individual simulations, specific tool use or predefined catalyst-screening protocols. However, a single study typically carries a high-level catalysis question through a chain of literature-derived structures, active-site choices, DFT and ML evidence, mechanistic interpretation and catalyst-design judgement across heterogeneous codes, stages and decisions. A key gap therefore remains for a research environment that preserves the scientific state of a catalysis investigation across fragmented software, long calculation chains and expert decisions while remaining anchored to scientific intent. Such an environment must connect model construction, DFT and ML calculations, mechanistic interpretation and catalyst-design judgement within a persistent, self-correcting research state, allowing autonomy to grow from high-level scientific intent instead of manual assembly of individual steps. It should also support the co-evolution of scientific state and modelling capability, so that validation feedback, expert guidance and reopened questions can refine the algorithms, routes and evidence used in subsequent steps, establishing a closed-loop autonomous agentic paradigm for materials design.

To address this challenge, we introduce CatMaster, an extensible multi-agent system that organizes specialist computational tools into a researcher-oriented agentic research system. Within a functionally complete runtime, CatMaster links executable atomistic modelling, adaptive algorithm refinement, internal critique within a persistent research state, allowing scientific questions, evidence standards and modelling routes to co-evolve during a computational investigation. We demonstrate CatMaster through a progressively demanding set of computational catalysis tasks ranging from standard executable scenarios and general-purpose materials modelling to literature-grounded mechanistic studies and a closed-loop design study on single-atom catalysts, a representative frontier catalyst family in which isolated metal centers and tunable coordination environments connect atomistic structure to activity and selectivity.[33,34] Across this ladder, CatMaster shows how natural-language intent can become calculation-ready models, atomistic evidence and catalyst-design conclusions within the same unified research ecosystem. Together, these studies outline a new computational catalysis paradigm in which human researchers focus on high-level scientific intents, while detailed validations for research questions are handled within a shared autonomous system for model setup, atomistic calculation, mechanistic analysis and catalyst-design decisions.

## Results

### From catalysis questions to autonomous executions

We began by testing whether CatMaster can carry routine computational catalysis tasks from a single natural-language request to usable scientific outputs. In each benchmark scenario, the researcher supplies a scientific task and final acceptance, while CatMaster sets up the workspace, chooses tools, prepares calculation inputs, executes the run, produces visual summaries and returns a short scientific report (Fig. 1a). This design measures whether the ecosystem can absorb routine operational work without close steering and provides the reliability baseline for the longer studies that follow.

The four benchmark scenarios cover core operations of computational catalysis practice. They include materials retrieval and DFT setup selection (T1), CO adsorption screening on Fe(110) (T2), a MACE-only transition-state benchmark on Au(111)-O (T3), and density-of-states (DOS) analysis for three bulk materials (T4) (Fig. 1b). Each scenario is scored only after the entire chain from workspace initialization to persisted scientific outputs is completed, including preparing inputs, executing calculations, generating visualizations and writing a summary report. This end-to-end requirement distinguishes our protocol from recent agentic studies that judge isolated tool invocations.[30] Important failure modes typically emerge only when reasoning, execution and reporting are combined into a complete scenario, and remain hidden in isolated protocol segments.[27] The benchmark therefore stresses the ecosystem interface as a complete operational unit.

Under this rigorous testing, four contemporary backends placed CatMaster into two clearly different operating tiers (Fig. 1c). GPT-5.4 achieved a perfect human-reviewed score of 100% across all four scenarios, closely followed by Sonnet-4.6 at 98%. Gemini-3.1 Pro reached 87.7%, and Mimo-V2 Pro (Hunter Alpha) scored 82.5%. Because long research cycles are built by compounding many short scenarios, even small per-scenario error rates accumulate rapidly along the chain. Near-ceiling reliability at the single-scenario level is therefore a prerequisite for long-horizon operation, and only GPT-5.4 and Sonnet-4.6 reach that bar in our evaluation. For a low-barrier catalysis ecosystem, this stratification has a concrete operational meaning. With the top-tier backends, the human user can hand off scenario-scale tasks and review the completed deliverables. With the weaker backends, the same chain still requires explicit human checkpoints on individual hand-offs.

The lower-scoring runs reveal backend-dependent reliability limits within the shared CatMaster runtime. Each backend dropped points for distinctly different reasons (Fig. 1d). Gemini-3.1 Pro produced chemically reasonable outputs but often failed to close a scenario end-to-end, missing final summaries or leaving computational tasks incomplete on T2 adsorption and T4 DOS. Mimo-V2 Pro showed much larger run-to-run variance and, on T4 DOS, suffered a near-catastrophic failure by incorrectly restarting an earlier step. Even the strongest backends might retain narrow blind spots when tracking deep mechanistic context. On T3, Sonnet-4.6 completed the nudged elastic band (NEB) calculation but performed the vibrational analysis on the endpoint minimum rather than on the transition-state image, accounting for its only consistent point loss. These backend-specific profiles form a reliability map of the shared runtime. Top-tier backends preserved closure across the request-to-report chain, supporting low-intervention setup, execution and reporting for routine scenario-scale tasks. Lower-tier backends lost reliability at distinct links, including scenario closure, run continuity and physically specified post-analysis, establishing where expert supervision remains necessary when the same runtime is paired with less reliable model backends.

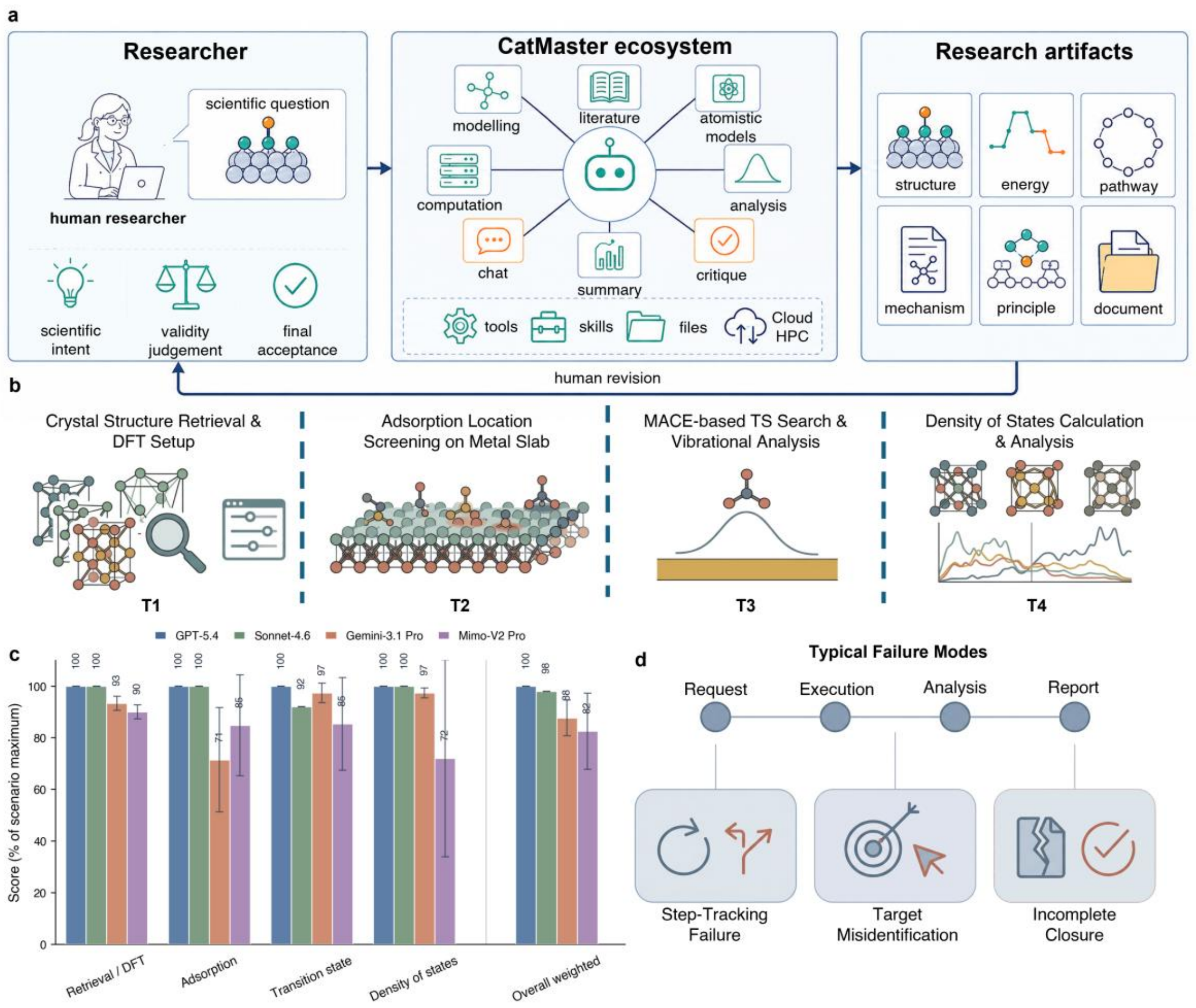

**Figure 1 | CatMaster as a virtual ecosystem for computational catalysis.** (**a**) Conceptual organization of the interface. The researcher provides the scientific question and later judges validity and final acceptance, while the CatMaster ecosystem connects modelling, literature, computation, atomistic models, analysis, critique and project files to persistent research artifacts. (**b**) Four routine benchmark scenarios used to test the interface, covering crystal-structure retrieval and DFT setup, adsorption screening, MACE-based transition-state search and density-of-states analysis. (**c**) Scenario-normalized scores for four model backends, reported as means over three independent repeats with error bars showing repeat standard deviations. (**d**) Failure modes observed across the benchmark chain, grouped by step tracking, target identification and scenario closure.

## Adaptive modelling strategies to materials data

We next asked whether the same ecosystem can adapt across heterogeneous materials data tasks, where the human user fixes the predictive target while CatMaster must independently choose the inductive bias. We tasked the system with constructing predictive models from scratch across six MatBench scenarios spanning composition-only and structure-aware inputs, regression and classification targets, and dataset sizes from 312 to 106,113 samples (Fig. 2a).[35] The selected scenarios were *steels*, *expt_gap*, *glass*, *jdft2d*, *mp_is_metal* and *phonons*. Each scenario started from the same natural-language request and required the runtime to perform feature engineering, model selection and MatBench-compliant held-out evaluation, with up to five internal strategy rounds. The agent therefore decided whether a given task favored descriptor-based tabular models, structure-aware handcrafted features, hybrid ensembles or graph neural networks (GNNs) (Fig. 2b).

Across the complete MatBench protocol, CatMaster reached top-ranked or near-leaderboard performance in all six scenarios with five scenarios reached this level in a single trial (Fig. 2c–g). CatMaster ranked first on *jdft2d* (33.09 meV atom$^{-1}$ MAE; Fig. 2d) and *mp_is_metal* (0.9787 ROC-AUC; Fig. 2g), and second on *steels* (80.66 MAE; Fig. 2e), *expt_gap* (0.3025 eV MAE; Fig. 2c) and *glass* (0.9524 ROC-AUC; Fig. 2f). These rankings were obtained as end-to-end agentic modelling runs, not through manual architecture selection or specialized AutoML search. Inspection of the saved workspaces shows a consistent task-conditioned pattern under the five-strategy budget. For composition-dominant scenarios (*steels*, *expt_gap*, *glass*), the system converged on descriptor-heavy tabular pipelines using ExtraTrees or XGBoost variants over matminer- or pymatgen-derived features.[6-8] For structure-dependent scenarios, the representation became broader. On *jdft2d*, CatMaster assembled composition, 2D geometry and CrystalNN-derived site statistics before selecting a weighted ExtraTrees-HistGradientBoosting model ensemble.[36] On *mp_is_metal*, it combined composition, elemental fractions, lightweight structure descriptors and neighbor statistics into a 322-feature XGBoost pipeline. These choices indicate that the runtime emulates the standard practice of a human data scientist by testing a finite set of inductive biases against the data properties, selecting the strongest candidate by robust cross-validation, and executing the resulting pipeline without the user having to specify the route in advance.

The *phonons* scenario, however, revealed a different operating mode of the ecosystem (Fig. 2h). Phonon-frequency prediction often requires geometry-rich, specialized architectures beyond the general-purpose representations that succeeded in the other tasks. In the first run, CatMaster selected a validation-weighted blend of a descriptor model and an underperforming crystal GNN inspired by CGCNN[37] and MEGNet[38], yielding 45.56 cm$^{-1}$ MAE. This performance is substantially weaker than the specialized phonon leaders and unlike the other five scenarios that reached near-leaderboard performance in a single pass. Inspection of the saved workspace suggested that the gap reflected a missing geometry-rich inductive bias, we therefore reinitiated the task to test self-improvement under minimal human oversight, only granting approval for further attempts without specifying architectures, features or validation methods. Across three iterations, CatMaster actively revised its ML modelling strategy and reduced the MAE from 45.56 to 37.69, 32.03 and finally 28.92 cm$^{-1}$, competitive with leading algorithms. For its final pass, the agent moved away from broad general graph neural network exploration that it tried at the beginning. Instead, the agent developed a physics-informed tabular route, combining robust descriptors with pseudo-dynamical spectral features derived from periodic mass-weighted spring graphs under conservative train-only selection, and reduced the error.

Together, these MatBench results demonstrate the dual operational modes of the agentic system. If standard modelling methods and features are sufficient familiar,

CatMaster rapidly combines and tunes them to reach near-leaderboard performance in a single trial. Conversely, when an initial strategy lacks the appropriate inductive bias, the system can also enter a self-improvement phase where benchmark feedback and historical artifacts drive capability evolution. Under this paradigm, the human expert shifts from generating models to evaluating them, by preventing premature acceptance, directing continued exploration and validating the scientific reasonability of the final output. This dynamic becomes especially critical in literature-grounded catalysis. In such scenarios, the challenge of missing inductive biases is no longer reflected by a scalar benchmark, but by the intricate resolution of three-dimensional structures, species identities and local reaction pathways that will be discussed in following sections.

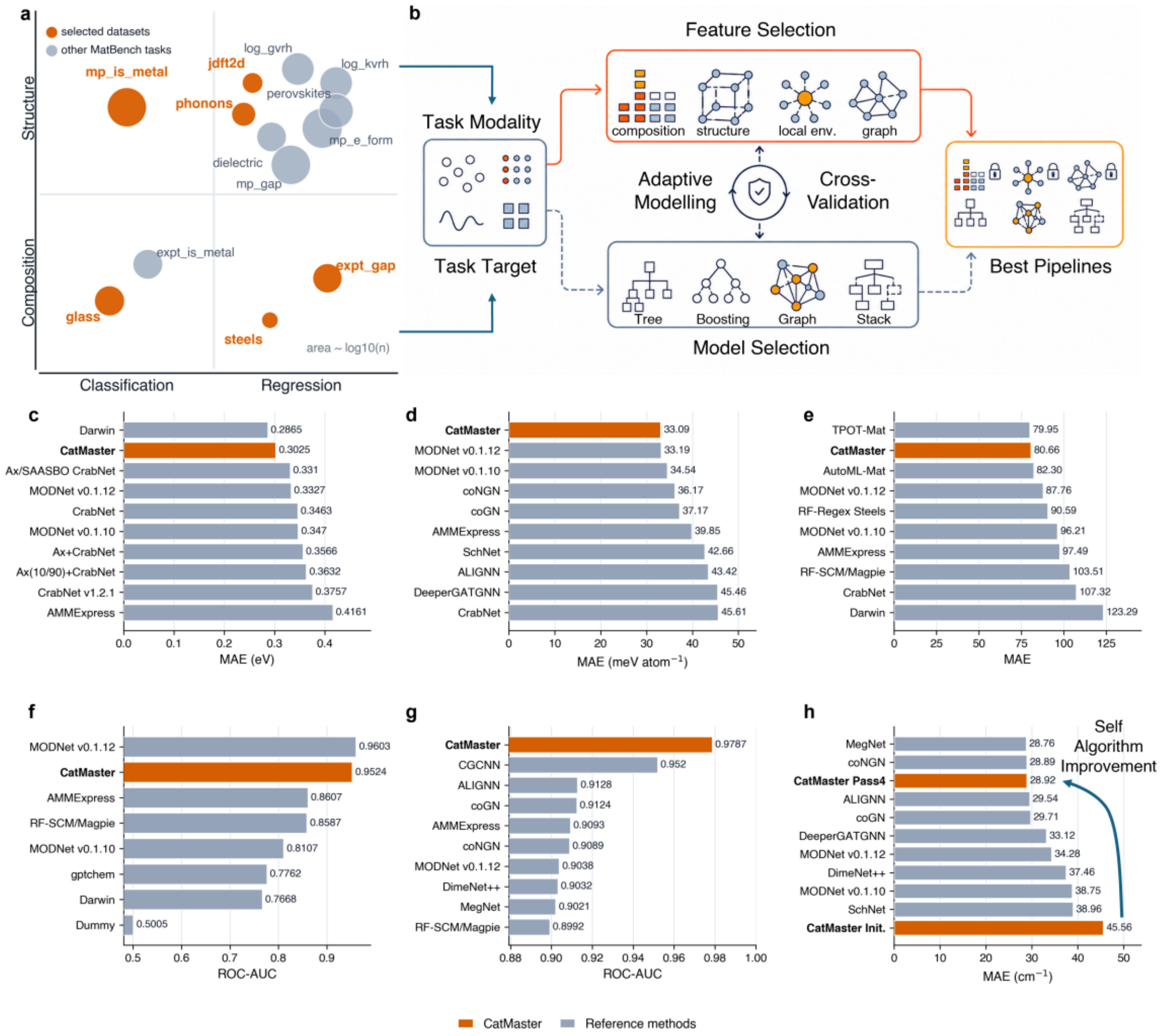

**Figure 2 | Adaptive materials modelling across the MatBench task space. (a)** Position of the selected MatBench datasets relative to the full benchmark set, organized by input primitive and prediction type. Circle area scales with $log_{10}(n)$, where $n$ is the number of samples, and orange marks the six datasets used here. **(b)** Schematic of the modelling procedure, in which task modality and task target guide feature-family selection, model selection and cross-validation before the best pipeline is retained. **c–h,** Leaderboard comparisons for *expt_gap* (**c**), *jdft2d* (**d**), *steels* (**e**), *glass* (**f**), *mp_is_metal* (**g**) and *phonons* (**h**). Orange bars denote CatMaster runs, and grey bars denote reference

leaderboard methods. Regression panels report MAE in the units shown; classification panels report ROC-AUC.

## Turning historic literature into calculation-ready models

Having established low-barrier operation and task-adaptive modelling in executable benchmarks, we next asked whether CatMaster could formulate atomistic models for catalyst systems drawn from active research fronts. Computational catalysis is often used to interpret or guide design in such areas, where a key step is to translate a catalyst precedent into an atomistic representation that preserves the intended active site, adsorbate state and reaction geometry. This step moves beyond standardized tasks, placing CatMaster at the model-construction stage of real catalytic research.

We therefore selected representative published catalysis studies and asked CatMaster to reconstruct the key computational models and calculation inputs from paper manuscripts. The four retained cases cover a spectrum of typical computational catalysis models, including oxide-supported Ru clusters for $NH_3$ activation[39], zeolite methylation endpoints[40], Cu/K/water electrochemical $CO_2$RR interfaces[41], and gradient PtFeCoNiCu high-entropy-alloy HER models[42]. Together, they span a spectrum of supported metal/oxide catalysis, confined zeolite chemistry, electrochemical interfacial catalysis and compositionally complex alloy electrocatalysis, sampling model-construction problems that recur across contemporary heterogeneous catalysis. This case series tests whether the same runtime can convert highly heterogeneous literature precedents into calculation-ready atomistic structures. The conversion is challenging because calculation-critical details, including surface terminations, adsorption registries, slab conventions, defect definitions and endpoint choices, are often split across figures, shorthand notation, supplementary details and tacit community knowledge.

Across these cases, CatMaster produced representative model sets that preserved the intended catalytic motifs in the initial run for all studies. The same tests also showed why a single-pass natural-language interaction remains constrained by the ambiguities of scientific literature. Papers often omit explicit facet-selection criteria, and current agents still struggle with species-aware geometric reasoning. Consequently, some initially generated states contained physical inaccuracies such as incorrect surface terminations or mismatched slab sizes. Following one to two rounds of explicit human clarification, CatMaster corrected these deviations and converged on the precise, calculation-ready geometries (Fig. 3b–e). For instance, in the Ru/$RE_2O_3$ $NH_3$-activation case (Fig. 3a), the human user rejected the initial surface interpretation and supplied a revised view of the vacancy, slab-size notation, and $\gamma$-$Al_2O_3$ termination. CatMaster then incorporated the correction and regenerated all corresponding calculation inputs.

These results show that in literature-grounded catalysis, the remaining human role shifts from routine construction toward semantic arbitration. CatMaster absorbed much of the labor-intensive translation from heterogeneous scientific descriptions to atomistic artifacts, while brief expert inputs remained necessary for under-specified physical choices that the original papers did not encode as executable instructions. The system can perform most construction and correction propagation, whereas ambiguous physical meaning in the source material still requires expert judgement.

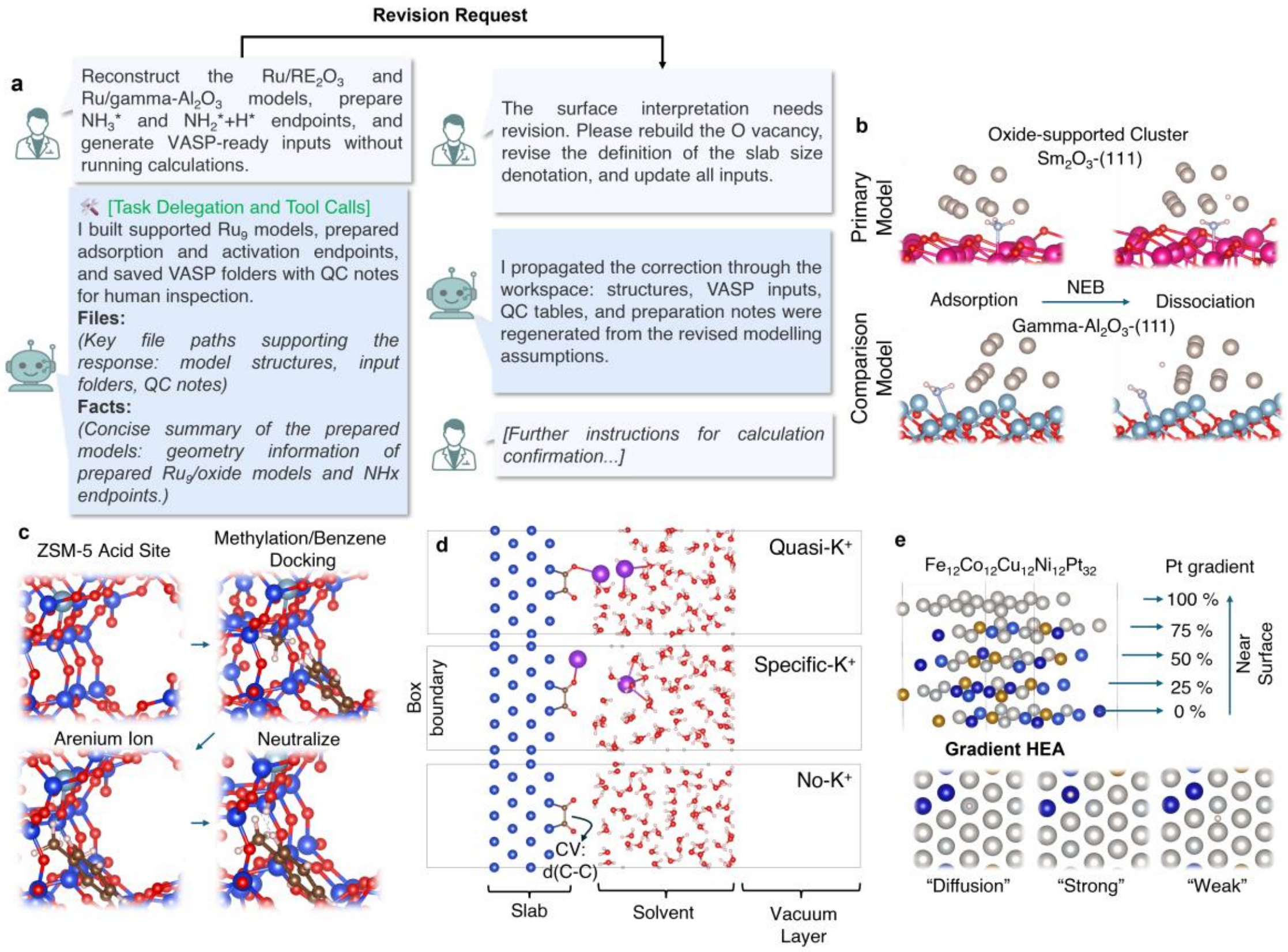

**Figure 3 | Literature-grounded construction of calculation-ready catalyst models.** **(a)** Ru/$RE_2O_3$ and Ru/γ-$Al_2O_3$ $NH_3$-activation case, showing the initial natural-language request, returned file and fact summary, human correction and propagation of the revised surface interpretation into structures, inputs and notes. **(b)** Corrected $Ru_9$/$Sm_2O_3$ and $Ru_9$/γ-$Al_2O_3$ endpoint models for $NH_3$ adsorption and first N–H activation. **(c)** Zeolite methylation structures and endpoint skeletons reconstructed from the published study. **(d)** Cu/K/water electrochemical-interface models for $CO_2$RR, comparing quasi-specific, specific and K-free *CO–*CO coupling seeds. **(e)** Gradient PtFeCoNiCu HEA(111) models for HER, including composition-controlled slabs, H* adsorption starts and spillover-path seeds. Complete workspaces are provided in Supplementary Data.

## Mechanistic exploration on catalytic surfaces

We next tested whether CatMaster could move from static model construction to mechanism-level exploration, where scientific validity depends on preserving chemically meaningful transformations across continuously changing states. A mechanism agent must maintain adsorbate identity, surface registry, local chemical correspondence and path locality while generating endpoint structures and transition-state searches. We therefore paired a compact rediscovery task, CO oxidation on Pt(111)[43,44], with a more complex RWGS network on Cu(111)[45,46]. For both systems, CatMaster was tasked with selecting a low-coverage slab, enumerating candidate minima, launching nudged elastic band (NEB) calculations, and synthesizing a mechanistic summary.[47] Because unconstrained agentic exploration of such highly complex, multifaceted reaction networks using full DFT is computationally prohibitive, we initially restricted the agent to operate using the MACE surrogate potential across both scenarios.

On the Pt(111) surface, CatMaster successfully recovered the canonical low-coverage mechanism. Operating on a 3×3×4 slab using MACE, it identified stable CO adsorption on atop sites, shallow molecular $O_2$ precursor states, a lower-energy dissociated O*+O* basin, and a direct O*+CO* oxidation event as the rate-controlling step. The computed energy barriers were 0.412 eV for O diffusion, 0.952 eV for $O_2$ dissociation, and 1.129 eV for direct oxidation. Following approximate thermodynamic corrections at 400 K, the direct oxidation step correctly remained the kinetic bottleneck with a 1.114 eV free-energy barrier (Fig. 4b). Although the system also explored an explicit molecular $O_2$-assisted branch (Fig. 4a), no competitive barrier survived its rigorous internal acceptance criteria. The system did more than reproduce a single expected value. It accurately reconstructed the accepted mechanistic hierarchy, confirming that dissociative $O_2$ activation feeds a lower-energy basin and that the Langmuir-Hinshelwood mechanism dictates the macroscopic rate.

The Cu(111) RWGS study was treated as a complex experiment and exposed a different challenge in a chemically dense and geometrically ambiguous reaction space. The agent repeatedly advanced, stalled and required redesign rather than showing a single-point failure. Under MACE, CatMaster expanded the low-coverage state space, recovered plausible local basins, and proposed a chemically meaningful network by algorithmic enumeration for local states, successfully isolating several locally supported segments, including a cleavage-like step of 0.364 eV. Yet most initial NEB calculations failed to converge, and attempts to extend these segments into a continuous RWGS network broke at one or more handoffs. An inspection showed that many NEB calculations stalled before clean convergence, trapped in non-decaying force plateaus under the MACE potential, whereas others relaxed into nearby basins (Supplementary Information) and left an unconverged reaction network (Fig. 4c).

To separate the agent's spatial-reasoning limitations from surrogate force-field errors, we directed the system to restart the case study using DFT as the backend. That follow-up recovered several short, chemically interpretable local connectors, including product-side steps with forward barriers of 0.222 eV between a local CO + H + OH product-equivalent minimum and a CO-shifted connector state, 0.135 eV between that CO-shifted connector and a neighboring free-H-shuffled CO + H + OH state, and 0.252 eV between an adjacent-H CO + H + OH intermediate and a split-intermediate pre-association state. At the same time, more than half of 43 intended NEB calculations failed to converge successfully and several problematic paths were already ill-posed at initialization. Manual and programmatic inspection of representative NEB inputs showed that nominally local transformations often failed to preserve coherent adsorbate identity across the interpolated images, especially in the $CH_2O_2$-related chemical space. In one case, a putative HCOO* migration evolved toward a CO + OH-like dissociation pattern. In another, an intended OCHO/HCOO rearrangement became entangled with spurious lateral translation. Taken together, these results locate the limitation in chemically local, species-aware path construction rather than in access to tools alone. In this regime, reliable mechanism discovery through a low-intervention interface requires stronger geometry-aware path construction and, ideally, specialized mechanistic frameworks such as CARE.[48]

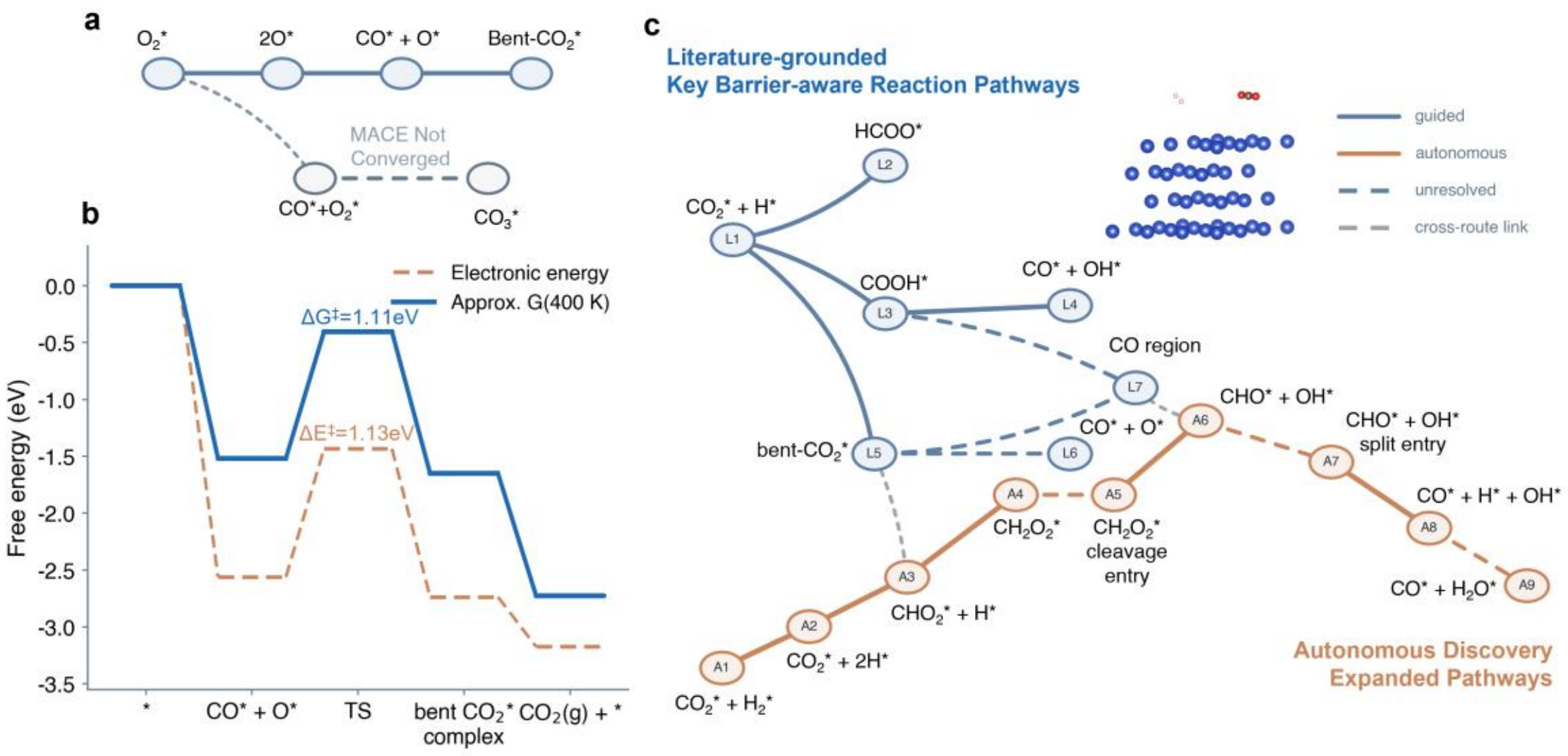

**Figure 4 | Mechanism rediscovery and reaction-network exploration. (a)** Pt(111) CO-oxidation network recovered from MACE searches, including the converged $O_2$ dissociation and O*+CO* oxidation route and an explored molecular $O_2$-assisted branch that did not converge. **(b)** Electronic-energy and approximate 400 K free-energy profiles for the recovered direct oxidation path, with the oxidation transition state remaining the highest barrier. **(c)** Cu(111) RWGS pathway map combining literature-guided reaction families with agent-expanded local connections. Solid blue and orange edges denote guided and expanded links, dashed edges mark unresolved joins, and grey dashed edges mark cross-route correspondences.

## Closed-loop single-atom catalyst design via self-feedback

Having moved from model construction to mechanism-level exploration, we next asked whether CatMaster could carry a catalysis question into prospective catalyst design. We chose single-atom catalysts (SACs) for this design study, as isolated metal atoms on tailored supports combine near-100% atom utilization with well-defined active centers and tunable activity and selectivity, making SACs a major direction in heterogeneous catalysis for $CO_2$ reduction, oxygen electrocatalysis, hydrogen evolution, nitrogen reduction and small-molecule activation.[33] These features also create a natural design space for computational catalysis, where metal identity, first-shell coordination and second-shell support chemistry can be varied systematically and ranked by stability and reaction free energies. Graphene-supported $MN_4$ motifs are widely used in synthesis and computational descriptor studies, allowing CatMaster to connect literature grounding, ML-DFT screening, thermodynamic ranking and mechanistic interpretation.

Starting from a single natural-language catalysis query, we challenged CatMaster to design graphene-supported $MN_4$ catalysts for two-electron $CO_2$ reduction to CO by varying the metal centre, first-shell N-site substitution and second-shell C-site substitution under a combined ML-DFT strategy. CatMaster grounded this brief in its own literature survey, using prior $MN_4$ $CO_2RR$ studies to define metal identity, first-shell coordination and neighbouring graphitic heteroatoms as established tuning handles.[49] Within this chemically constrained design space, the system selected concrete metals and dopants and organized them into a shell-resolved motif matrix.

CatMaster first ran a dispersion-inclusive MACE mh-1 survey of 29 motifs on a graphene-supported $MN_4$ template, covering bare, COOH* and CO* states together with finalist $CO_2$* and H* checks (107 total relaxations) under a vacuum computational hydrogen electrode (CHE) model.[50] The screening report identified second-shell P as the strongest tested perturbation for lowering the COOH* formation descriptor and selected P–$CoN_4$ and P–$NiN_4$ for follow-up (Fig. 5a,b). At this stage, the research controller could already assemble a coherent screening report, creating a premature-closure risk common in agentic research systems. The internal review specialist was designed to interrupt that path. It recast the P-centred trend as a validation hypothesis and returned missing evidence checks to the research controller, noting that the ranking depended on a manuscript-defined score, softened a large CO-release penalty, lacked DFT calibration and used electronic energies only.

CatMaster then produced a four-motif VASP+D3 calibration set across bare, COOH* and CO* states (Fig. 5c). The second review accepted the safer framing of P motifs as validation-priority candidates, yet the calibration data also exposed a deeper evidence gap. MACE-to-VASP shifts could reach 1–2 eV for key energy descriptors, making fine ranking from the original surrogate screen unreliable. Addressing this gap required a larger computational campaign with free-energy corrections, solvation treatment and broader surrogate validation. Within the initial authorization, CatMaster revised the research package by clarifying the DFT method, treating the VASP subset as calibration evidence, discussing the surrogate discrepancy and narrowing the conclusion to an electronic-energy activation-versus-release tradeoff. The third review still judged the project below scientific evidence standard, and the researcher then approved expanded computation.

CatMaster next built a 35-motif DFT matrix spanning pristine, single first-shell and single second-shell variants across bare, COOH* and CO* states, while repairing the local surrogate model and strengthening the thermodynamic basis of the ranking (Fig. 5d,e). The expanded matrix moved the design picture away from the original P-led ranking. P–$CoN_4$ retained favorable early-step energetics, yet its activation gain was coupled to a large CO-release penalty, making it an overbinding comparator. The DFT matrix carried forward B–$CoN_4$, $NiN_3B$, $CoN_4$ and N–$CoN_4$ as the main candidate set, while Cu $N_4$ and P–$CoN_4$ defined the under-activation and overbinding boundaries. CatMaster aligned MACE on the accumulated DFT data, reducing the energy MAE from 149.75 to 11.72 meV $atom^{-1}$ and the force RMSE from 0.109 to 0.027 eV $Å^{-1}$. The aligned model supported follow-up prioritization within the Co/Ni $N_3B$ neighbourhood, nominating N–$NiN_3B$ for DFT confirmation and rejecting B–$NiN_3B$ because of severe hydrogen overbinding. In parallel, CatMaster added targeted implicit-solvation validation and explicit adsorbate-frequency calculations. Together, these additions established the final ranking. The three later internal reviews found that the main scientific problems had been resolved while the scholarly presentation still needed refinement. Guided by these critiques, CatMaster improved how the study framed its evidence, claim scope and remaining mechanistic limits.

Under the final solvated CHE ranking, B–$CoN_4$, $NiN_3B$ and N–$NiN_3B$ emerged as the optimal motifs, with limiting proton-coupled electron transfer costs of 0.015, 0.062 and 0.080 eV, respectively; a recent peer-reviewed theoretical investigation independently highlights $NiN_3B$ (Fig. 5f–h).[51] The internal review sequence therefore documents a scientific correction across the design trace. The initial surrogate screen pointed to a P-centered activation trend, then DFT arbitration and thermodynamic treatment instead supported a boron- and nickel-centered design conclusion, with the phosphorus branch retained as the over-binding limit of the catalyst design space and supported the final recommendations.

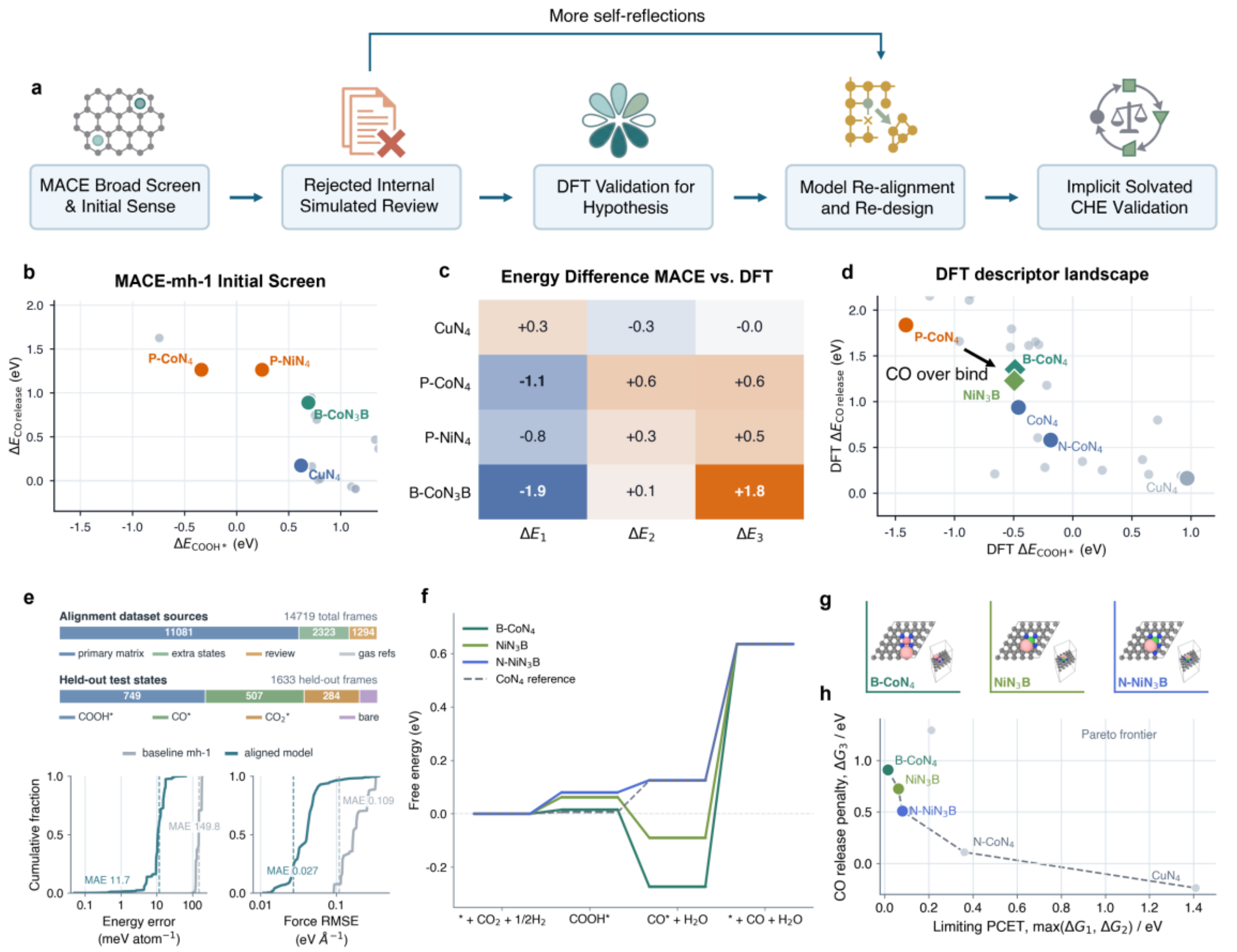

**Figure 5 | Single-atom catalyst design through self-feedback. (a)** Design trajectory for the graphene-supported $MN_4$ $CO_2$-to-CO study, from MACE broad screening and internal review to DFT validation, local model re-alignment and solvated-CHE validation. **(b)** Initial MACE descriptor map for COOH* formation and CO release, highlighting the P-centered candidates and calibration contrasts selected for follow-up. **(c)** VASP+D3 minus MACE descriptor shifts for the four calibration motifs, showing large surrogate errors for key descriptors. **(d)** Expanded DFT descriptor landscape for pristine, first-shell and second-shell motifs, in which P–$CoN_4$ becomes an overbinding comparator and B–$CoN_4$, $NiN_3B$, $CoN_4$ and N–$CoN_4$ remain as design candidates. **(e)** Alignment-set composition and held-out error distributions after local MACE re-

alignment. **(f)** Solvated-CHE free-energy profiles for the $CoN_4$ reference and promoted motifs. **(g)** Representative optimized structures for the three final motifs. **(h)** Final activation–release landscape, where B–$CoN_4$, $NiN_3B$ and N–$NiN_3B$ occupy the low-limiting-PCET and low-CO-release region.

## Discussion

Taken together, these results establish a new operating paradigm for autonomous computational catalysis, in which high-level catalytic intent is carried through a persistent, self-correcting virtual ecosystem of atomistic models, computational evidence, mechanistic analysis and catalyst-design judgement. In this environment, a natural-language catalysis question can remain connected to atomistic modelling, mechanism analysis and catalyst-design decisions throughout the study. Across the studies reported here, CatMaster progressed from routine executable scenarios to task-adaptive materials modelling, literature-grounded model construction, mechanism exploration and a SAC campaign that revised an initial P-centered trend into a B- and Ni-centered design conclusion. Across these cases, self-evolution occurred at two coupled levels. The research state evolved from hypotheses and provisional claims into supported mechanistic or design conclusions, while the capability state evolved through new scripts, revised modelling routes, and locally calibrated surrogate models.

Specifically, MatBench evaluations reveal two distinct modes of agentic modelling. For five scenarios, established feature families and model classes were sufficient for CatMaster to reach top-ranked or near-leaderboard performance in one trial. The phonon dataset presented a different challenge, as the initial run lacked a geometry-rich inductive bias. Yet, three self-improvement cycles transformed the original underperformance model into a near-leading result. Relying on its own logic instead of user-specified design, CatMaster generated a physics-informed pseudo-dynamical spectral representation through self-experiments, and demonstrates that modelling capabilities can successfully evolve together a persistent scientific state.

However, the mechanistic case studies identify a boundary in current geometric reasoning for LLM-based agentic systems, which cannot be captured by a single scalar loss and expert arbitration is still required for adsorbate identity, local path correspondence, and surface registry. On Pt(111), CatMaster recovered the accepted low-coverage CO-oxidation hierarchy. On Cu(111), the same runtime could expand the RWGS state space and recover several local connectors, yet many proposed NEB tasks became fragile once adsorbate identity and path locality had to be preserved across a dense surface network. Several nominally local paths mixed the intended chemistry with fragment drift or lateral migration before electronic-structure accuracy became the decisive issue. Surface catalysis papers also encode adsorption sites, slab registries, intermediate identities and local reaction motifs through shorthand notation or figures, leaving ambiguities that a general text-and-code agent cannot reliably resolve alone.

This difficulty reflects a mismatch between code-mediated and visual geometric reasoning. Human modelers often recognize implausible placements, broken correspondences or unintended adsorbate motion by inspecting the three-dimensional scene. Agents express the same operations through coordinates, atom indices, constraints and file transformations, where small mistakes can silently create corrupted geometries or mismatched inputs. This concern is consistent with recent observations that even strong coding agents have limited reliability when reproducing computational materials-science findings from original paper materials.[53] Physical validation is also difficult to reduce to deterministic software tests, because the central question is often whether a structure, path or surface model preserves the intended chemistry. These checks still rely on tacit domain judgement.

These boundaries clarify the evolving role of human scientists in agentic computational catalysis. In our study, the researcher specified high-level intent, authorized resource expansion, resolved literature ambiguities and set the evidence standard for final claims. CatMaster provided scientific critique, revised the evidence hierarchy and executed follow-up calculations once the study was reopened. This division of labor concentrates human effort on research framing, ambiguity arbitration and evidence-standard decisions.

Looking forward, scientific agents may follow scaling relations that differ from neural scaling laws, where loss scales with model size, data and compute.[54,55] For scientific agents, performance is likely governed by a coupled relation among foundation-model capability, tool coverage, verifiable feedback, domain-specific validators and the quality of human-agent correction traces. Recent work suggests that coordination gains in agent systems depend on task structure and architecture, with multi-agent designs benefiting decomposable tasks and degrading sequential or tool-heavy ones.[56] Studies of self-improving coding agents and self-adapting language models further show that closed-loop improvement is plausible in constrained environments with benchmarks or reward signals.[57] In computational catalysis, the decisive scaling variables are physically grounded model construction, species-aware geometric reasoning, reaction-path validation and evidence-standard arbitration. Human-agent research trajectories may therefore become the most valuable training substrate. Expert corrections, failed calculations, repaired structures, critique outcomes and resource decisions record where agent reasoning meets catalytic reality. Reusing these traces could guide future agentic reinforcement learning and the co-evolution of computational-catalysis agents with expert physical judgement.

## Methods

### Agent runtime

CatMaster is organized around three method-level requirements of open-ended computational catalysis. A study must preserve research-state continuity across long calculations, transform structures, calculations, datasets and models through a composable atomic tool surface, and connect execution with interpretation and revision through reusable scientific procedures. The implementation follows these requirements through a shallow specialist hierarchy, an atomic tool surface, a decoupled skill layer, workspace-scoped execution tools, structured reporting and internal review. The full runtime architecture, including specialist roles, the tool surface and the underlying execution infrastructure, is shown in Supplementary Fig. S1.

These requirements can be summarized as a persistent research state,

$$\mathcal{D}_t = (q, \mathcal{H}_t, \mathcal{O}_t, \mathcal{C}_t),$$

where $q$ is the catalysis question, $\mathcal{H}_t$ contains the current model assumptions and design hypotheses, $\mathcal{O}_t$ is the accumulated computational record, and $\mathcal{C}_t$ is the provisional mechanistic or design conclusion drawn from $q$, $\mathcal{H}_t$ and $\mathcal{O}_t$. The record is a set of artifact instances,

$$\mathcal{O}_t = \{a_i\}_{i=1}^{n_t}, \qquad \mathrm{type}(a_i) \in \mathcal{A},$$

where $\mathcal{A}$ is the artifact vocabulary defined below. The research specialist maintains $\mathcal{D}_t$, the tool surface expands or revises $\mathcal{O}_t$, and the skill and review layers govern the transition from $\mathcal{D}_t$ to the next research state or to a scientific stopping point.

The first requirement is research-state continuity. Computational catalysis studies often pass through failed relaxations, negative screening trends, resource decisions and reopened mechanistic questions. CatMaster addresses this through a shallow hierarchical architecture organized for context isolation and long-horizon continuity using Langchain deepagents. A central design choice is to keep the hierarchy shallow to minimize context loss and as broad as the scientific work demands. This keeps high-level reasoning close to the workers that manipulate structures, calculations and analysis records, reducing hand-offs that can scatter intermediate state across many controllers.

At the apex of this hierarchy is the *Research Specialist*, the sole orchestration-capable controller responsible for maintaining the active research state. It delegates focused tasks to four primary mid-level coordinators. The *Experiment Specialist* routes computational materials execution, machine learning, molecular quantum chemistry and structured result reporting to dedicated workers. The *Writing Specialist* organizes accepted evidence into scholarly reports and presentation-ready scientific text, with claim strength tied to the supplied calculations and citations. The *Literature-review Specialist* connects broad literature search with verified scholarly metadata. The *Peer-review Specialist* provides internal scientific critique by reading the same evidence package or compiled report and identifying weak mechanistic support, thermodynamic gaps, missing validation or unclear scientific presentation before the study proceeds.

By confining execution- or writing-heavy episodes to these dedicated domain specialists and their respective workers, CatMaster reduces delegation depth. Each execution episode maintains persistence across conversation turns through stable thread identifiers and workspace-level state stores. This persistence carries computational state across human interventions, so that a study can pause for arbitration or authorization and resume on the same artifact tree without restating the full history of calculations, intermediate files and prior decisions in the active model context. Unless otherwise stated, all non-benchmark experiments reported after the model-comparison benchmark were executed with GPT-5.4 using high reasoning effort, selected as the strongest backend from the benchmark evaluation.

## Agent tool interface

The second requirement is a composable atomic tool surface. CatMaster implements this surface so that every executable operation reads and writes typed scientific artifacts. Formally, each tool maps one artifact instance to another,

$$a_{\text{in}} \xrightarrow{T_k} a_{\text{out}}, \qquad \text{type}(a_{\text{in}}), \text{type}(a_{\text{out}}) \in \mathcal{A},$$

where $\mathcal{A}$ denotes the typed artifact vocabulary used by the runtime,

$$\mathcal{A} = \{S, SB, C, CB, R, RB, A, D, M\}.$$

Here $S$ and $SB$ are single structures and structure batches, $C$ and $CB$ are calculation-input directories and input batches, $R$ and $RB$ are run-result directories and result batches, $A$ is an analysis artifact, $D$ is a dataset, and $M$ is a model. For an accepted tool output, the computational record is updated as

$$\mathcal{O}_t^{+} = \mathcal{O}_t \cup \{a_{\text{out}}\}.$$

A tool changes one artifact class at a time, including atomic structures, engine-specific input and output directories for VASP[58,59], ORCA[60,61], MACE[9], CP2K[62] and LAMMPS[63] workflows, datasets, trained models, and parsed analysis outputs. This primitive boundary lets the agent compose preparation, execution and analysis around the same scientific objects, while reasoning trajectory progression can be localized to the structure, input directory, parsed output, dataset or trained model that introduced them. In practice, the tool surface spans atomistic structure generation, batch execution, post-processing, literature retrieval, report compilation and machine-learning operations such as dataset construction and active-learning selection. The full inventory is broad at the system level, whereas each specialist sees a narrower view matched to its artifact classes and scientific responsibility.

The same interface also provides a self-adaptive extension path for operations beyond the initially registered tool set. When a required operation is absent, CatMaster can inspect the source of related built-in tools and reuse their input-schema, workspace and artifact-return conventions as implementation templates. It can then create persistent task-specific scripts in the active workspace's scripts directory, where they remain available for reruns, inspection and later reuse within the same project. The new code is stored as a persistent research artifact, rerun against the relevant structures, calculations or datasets, and checked before it contributes to a scientific conclusion. This mechanism is one route of modelling-capability self-evolution by enabling an operation that was unavailable at the beginning of a study to become a reusable artifact transformation after the agent implements, tests and stores it.

## Skills and workflow executions

Building upon composable tools, the third requirement is reusable scientific procedures that connect calculation execution with interpretation and revision. Structures, energies and model scores have to be interpreted in relation to the catalyst question that motivated them. This interpretation determines whether a mechanistic or design claim is sufficiently supported, or whether the study should re-enter calculation, analysis or model refinement, or called workflows in theoretical studies. CatMaster implements this requirement by linking a decoupled skill layer, workspace-scoped execution tools, structured reporting and internal review into a unified research loop.

State progression is governed by the active skill or review procedure $\sigma_t$,

$$\mathcal{D}_{t+1} = U_{\sigma_t}(q, \mathcal{H}_t, \mathcal{O}_t^+, \mathcal{C}_t).$$

The update operator $U_{\sigma_t}$ may preserve the expanded record, revise assumptions in $\mathcal{H}_t$, sharpen the provisional conclusion $\mathcal{C}_t$, or return a new required artifact type to the tool surface. The sequence reaches a scientific stopping point $\mathcal{D}_*$ when internal review finds that the accumulated computational record supports the mechanistic or design conclusion under the relevant skill-defined standards.

The skill layer supplies these procedures during this process. *Skills* describe domain practices such as slab construction, adsorption screening, transition-state routing and MACE dataset curation, and provide procedural guidance to specialists. Executable capability remains controlled by the active tool binding. This separation keeps procedure-level scientific choices editable during a study while runtime code remains stable. Only role-compatible skills are bound to any given specialist, and each skill composes the primitive tools available to that role. Detailed role, tool and skill inventories are provided in the Supplementary Information.

The execution surface supplies the calculation records used for interpretation and revision. Workspace-scoped tools handle structure manipulation, input generation, local analysis and external computational jobs. These tools create calculation directories, run or dispatch computations, parse outputs and return file paths or analysis records for downstream use.

Structured reporting and internal critique close the loop from calculation output to scientific judgement. After a calculation campaign reaches a provisional answer, the review lane examines whether the available structures, energies, model diagnostics and literature context support the intended mechanistic or design conclusion. The writing lane prepares scholarly reports from accepted structures, calculations, models and citations, with claim strength tied to the supplied materials. In the SAC study, this loop prevented the initial screening trend from becoming the final design conclusion and redirected the project, after human approval of a follow-up phase, toward stronger thermodynamic and mechanistic support. In this design, reporting and critique function as scientific self-correction within the computational study, connecting calculation outputs to interpretation, design judgement and follow-up computation.

## Computational scenario benchmark

Benchmark evaluations were launched through the same CatMaster experiment runtime used elsewhere in this study, with each run executed in an isolated workspace and initiated from a natural-language scenario request. We defined four benchmark scenarios that probe complementary aspects of computational catalysis practice. They cover structure retrieval and relaxation-input preparation, adsorption-energy screening, MACE transition-state analysis, and density-of-states interpretation. All models participating in the benchmark used high reasoning effort. For each model backend we ran three independent repeats per scenario, yielding 12 benchmark runs per model. Each candidate model had to complete the full chain from request to persisted scientific deliverables and a final report, ensuring that the benchmark evaluated the request-to-deliverable interface as a complete scenario.

Scoring was performed using LLM-as-a-judge[64] and a human-review check based on filesystem evidence using scenario-specific rubrics. During scoring, model backend identities were blinded and exposed only as anonymized labels from *Model A* to *Model D* to reduce potential model-name bias. Maximum scores and hard-cap rules were encoded in the benchmark judge. The judge inspected only persisted outputs under each run workspace, rewarding equivalent evidence even when local file layouts differed. The human-reviewed scores reported in Fig. 1c and discussed in the Results are the final benchmark values used in this manuscript; full prompts, rubric details, manual revisions and repeat-level score tables are provided in the Supplementary Information.

## MatBench modelling evaluation

General-purpose modelling evaluations were carried out on six MatBench[35] scenarios chosen to span task type, input primitive, and data scale. The selected scenarios were *steels*, *expt_gap*, *glass*, *jdft2d*, *mp_is_metal*, and *phonons*. Each scenario was launched from a common natural-language prompt template that asked CatMaster to complete feature engineering, model development, hyperparameter search, official benchmark evaluation, and report writing inside the local workspace. The prompt explicitly permitted up to five internal optimization rounds while preserving official MatBench data-splitting rules. For regression scenarios CatMaster used shuffled five-fold KFold splits with random state 18012019; for classification scenarios it used shuffled five-fold StratifiedKFold splits with the same seed using scikit-learn[65]. In all cases outer test folds were reserved for final reporting only, and feature, model, and hyperparameter selection were performed only within each outer-training partition.

For a modelling episode to count as complete, CatMaster had to produce reproducible scripts, saved models and held-out predictions, and a detailed markdown report in addition to the final metric summary. No manual code editing was inserted between the initial request and the official evaluation of any given scenario. For the *phonons* scenario, three explicit self-improvement episodes were authorized after the first run. Performance was compared against the MatBench results leaderboard as accessed in March 2026. The common prompt template, scenario matrix, leaderboard screenshots, recovered run summaries and full performance table are provided in the Supplementary Information.

## Single-atom catalyst design protocol

The SAC-design case began from a natural-language request to explore graphene-supported $MN_4$ catalysts for two-electron $CO_2$ reduction to CO under a limited exploratory budget. CatMaster started from a provided graphene/$FeN_4$ template structure and was required to choose an evidence hierarchy appropriate for catalyst design. It first used surrogate-assisted exploration to map doped $MN_4$ candidates and then escalated the emerging design hypotheses to DFT validation. After internal critique judged the first evidence package insufficient for a design conclusion, an explicit resource-authorization prompt opened a second phase focused on strengthening the thermodynamic and mechanistic basis of the final recommendation.

## Acknowledgements

This work was supported by the National Key R&D Program of China (No. 2022ZD0117501), the Scientific Research Innovation Capability Support Project for Young Faculty (ZYGXQNJSKYCXNLZCXM-E7), the Tsinghua University Initiative Scientific Research Program, and the Carbon Neutrality and Energy System Transformation (CNEST) Program led by Tsinghua University.

## Data availability

The relevant original workspaces required to reproduce this study are deposited on figshare with DOI 10.6084/m9.figshare.31891201

## Code availability

The source code required to reproduce this work is available at https://github.com/q734738781/CatMaster.

## Competing interests

The authors declare no competing interests.